\documentclass[12pt]{article}
\newlength{\pubnumber} 

\catcode`\@=11
\@addtoreset{equation}{section}

\def\section{\@startsection{section}{1}{\z@}{3.5ex plus 1ex minus .2ex}
 {2.3ex plus .2ex}{\large\bf}}
\def\subsection{\@startsection{subsection}{2}{\z@}{2.3ex plus .2ex}
 {2.3ex plus .2ex}{\bf}}

 \usepackage{graphicx}
\begin{document}

\begin{titlepage}
\samepage{
\setcounter{page}{0}
\rightline{August 2008}
\vfill
\begin{center}
    {\Large \bf  
  Correlation Classes on the Landscape:\\
   To What Extent is String Theory Predictive?\\}
\vfill
\vspace{.10in}
   {\large
      Keith R. Dienes$^1$\footnote{E-mail address:  dienes@physics.arizona.edu},$\,$ 
    Michael Lennek$^2$\footnote{E-mail address:  mlennek@cpht.polytechnique.fr}}
\vspace{.10in}

 {\it  $^1\,$Department of Physics, University of Arizona, Tucson, AZ  85721  USA\\}
 {\it  $^2\,$Centre de Physique Th\'eorique, Ecole Polytechnique, \\
            CNRS, F-91128 Palaiseau, France\\}
\end{center}
\vfill
\begin{abstract}
  {\rm   In light of recent discussions of the string landscape,
       it is essential to understand the degree to which string theory
       is predictive.  We argue that it is unlikely that the landscape as a whole
       will exhibit unique correlations amongst low-energy observables, but rather
       that different regions of the landscape will exhibit different overlapping
       sets of correlations.  We then provide a statistical method 
       for quantifying this degree of predictivity, and for extracting
       statistical information concerning the relative sizes and overlaps
       of the regions corresponding to these different correlation classes.
       Our method is robust and requires no prior knowledge of landscape
       properties, and can be applied to the landscape as a whole as
       well as to any relevant subset.
      }
\end{abstract}
\vfill
\smallskip}
\end{titlepage}

\setcounter{footnote}{0}

\def\beq{\begin{equation}}
\def\eeq{\end{equation}}
\def\beqn{\begin{eqnarray}}
\def\eeqn{\end{eqnarray}}
\def\half{{\textstyle{1\over 2}}}

\def\calO{{\cal O}}
\def\calE{{\cal E}}
\def\calT{{\cal T}}
\def\calM{{\cal M}}
\def\calF{{\cal F}}
\def\calY{{\cal Y}}
\def\calV{{\cal V}}
\def\calN{{\cal N}}
\def\ibar{{\overline{\i}}}
\def\qbar{{\overline{q}}}
\def\mm{{\tilde m}}
\def\ahat{{\hat a}}
\def\nn{{\tilde n}}
\def\rep#1{{\bf {#1}}}
\def\ie{{\it i.e.}\/}
\def\eg{{\it e.g.}\/}

\def\boxit#1{\vbox{\hrule\hbox{\vrule\kern3pt
\vbox{\kern3pt#1\kern3pt}\kern3pt\vrule}\hrule}}

\def\Str{{{\rm Str}\,}}
\def\bone{{\bf 1}}

\def\thetai{{\vartheta_i}}
\def\thetaj{{\vartheta_j}}
\def\thetak{{\vartheta_k}}
\def\thetaibar{\overline{\vartheta_i}}
\def\thetajbar{\overline{\vartheta_j}}
\def\thetakbar{\overline{\vartheta_k}}
\def\etainv{{\overline{\eta}}}

\def\modinvmeasure{{  {{{\rm d}^2\tau}\over{\tautwo^2} }}}
\def\qbar{{  \overline{q} }}
\def\ahat{{ \hat a }}

\newcommand{\newc}{\newcommand}
\newc{\gsim}{\lower.7ex\hbox{$\;\stackrel{\textstyle>}{\sim}\;$}}
\newc{\lsim}{\lower.7ex\hbox{$\;\stackrel{\textstyle<}{\sim}\;$}}

\hyphenation{su-per-sym-met-ric non-su-per-sym-met-ric}
\hyphenation{space-time-super-sym-met-ric}
\hyphenation{mod-u-lar mod-u-lar--in-var-i-ant}


\def\inbar{\,\vrule height1.5ex width.4pt depth0pt}

\def\IC{\relax\hbox{$\inbar\kern-.3em{\rm C}$}}
\def\IQ{\relax\hbox{$\inbar\kern-.3em{\rm Q}$}}
\def\IR{\relax{\rm I\kern-.18em R}}
 \font\cmss=cmss10 \font\cmsss=cmss10 at 7pt
\def\IZ{\relax\ifmmode\mathchoice
 {\hbox{\cmss Z\kern-.4em Z}}{\hbox{\cmss Z\kern-.4em Z}}
 {\lower.9pt\hbox{\cmsss Z\kern-.4em Z}}
 {\lower1.2pt\hbox{\cmsss Z\kern-.4em Z}}\else{\cmss Z\kern-.4em Z}\fi}

\long\def\@caption#1[#2]#3{\par\addcontentsline{\csname
  ext@#1\endcsname}{#1}{\protect\numberline{\csname
  the#1\endcsname}{\ignorespaces #2}}\begingroup
    \small
    \@parboxrestore
    \@makecaption{\csname fnum@#1\endcsname}{\ignorespaces #3}\par
  \endgroup}
\catcode`@=12

\input epsf

\setcounter{footnote}{0}

Over the past few years,
the existence and implications of a vast string theory ``landscape''
have attracted considerable attention~\cite{landscape}. 
Indeed, research in this area has spanned a considerable range of topics
and followed a number of different 
approaches~[2--16];  
for recent reviews, see Ref.~\cite{reviews}.
However, because the specific
low-energy phenomenology that can be expected to emerge from string theory depends
critically on the particular choice of vacuum state within the landscape,
and because the space of possible string vacua is extremely large (with 
some estimates putting the number of phenomenologically interesting vacua at 
$10^{500}$ or more~\cite{douglas}),
the question which naturally arises is a critical one.
To what extent can we say that string theory is predictive?
In what sense can we say that certain low-energy phenomenological features
of the observed universe
are predicted by, or derivable from, string theory? 

The question of predictivity goes to the heart of what it means
to be doing science rather than mathematics.
As such, there can be no more critical question for string theory
than this.
Of course, predictivity is not an absolute necessity for 
all aspects of science --- 
indeed, good science often begins with observation and classification.
However, while observers and experimentalists need not be primarily concerned with
making predictions, theorists must be:
theories of science must incorporate the ability not only to explain, 
but also to predict. 
This is especially true for string theory, which,
as a branch of high-energy physics, 
must be judged by the prevailing standards of the field.
Moreover, even though many of the direct experimental consequences 
of string theory lie at 
presently inaccessible energy scales, not all will be.
And even if all of the firm experimental consequences of string
theory were somehow proven to lie at scales exceeding 
those reachable by 
current accelerator technology, this would not free string
theory from its obligations to make predictions which are testable 
at those higher energy scales ---
\ie, testable in principle, if not in practice.  

On the one hand, even accepting this standard,
one might argue that it is too much to ask that
string theory be predictive in and of itself.
From this perspective, one should rightly compare string theory not with a specific
quantum field-theoretic model such as the Standard Model, but
with quantum field theory itself --- indeed, both string theory and quantum
field theory can be viewed as languages or frameworks within which the subsequent act of
model-building takes place.  Just as the Lagrangian of the Standard Model is just one out of many
possible self-consistent quantum field-theoretic Lagrangians,    
the correct string model might be just one out of many possible
self-consistent string vacua.
Thus, according to this argument,  string theory is just as predictive
as quantum field theory:  neither becomes predictive until a particular
model is constructed, and all predictions that ensue can be expected to
hold only within that model.

While this argument has some validity, one could just as well argue that
it misses a critical point.  While quantum field theory tolerates many
free parameters, string theory does not:  generally all free parameters in
string theory (such as gauge couplings, Yukawa couplings, and so forth) are
determined by the vacuum expectation values of scalar fields and thus
are expected to have dynamical origins within the theory itself.
Moreover, while many architectural details of a given 
model (such as the gauge group, the number of generations, or even the 
degree of supersymmetry) are uncorrelated within quantum field theory,
string theory has deeper underpinnings in terms of the 
geometric properties and configurations of strings and branes.  
It therefore becomes meaningful to ask more from string theory than
from quantum field theory. 

Given the existence of the landscape,
it is certainly too much to demand that string theory give rise 
to predictions for such individual quantities as the number of 
particle generations.
Indeed, we already know that such individual quantities
can vary greatly from one string vacuum to the next.
However, it is perhaps not too much to ask that string theory 
manifest its predictive power through the existence of {\it correlations}\/
between physical observables that would otherwise be uncorrelated in quantum
field theory.
Such correlations would be the spacetime phenomenological manifestations
of the deeper underlying geometric structure that ultimately defines string
theory and distinguishes it from a theory whose fundamental degrees of freedom
are based on point particles. 
Of course, it is logically possible that string theory
leads to sharp correlations amongst observables at {\it high}\/ energy scales,
but that the mathematical form of the connections between these high-scale observables
and experimentally accessible low-scale observables completely washes
these correlations away as far as a low-energy physicist might be concerned.
However, there is no evidence that Nature is so cruel for the low-energy
parameters of interest.
Thus, our question concerning the predictivity of string theory 
boils down to a single critical question:
to what extent are there correlations between different physical observables 
on the string-theory landscape?

Clearly, the existence of such correlations across the string theory 
landscape would imply that string theory is
predictive, while the absence of such correlations would suggest that it is not.
Indeed, many recent discussions of this issue have proceeded under the assumption
that these are the only two logical options.

\begin{figure}[ht]
\centerline{
   \epsfxsize 4.0 truein \epsfbox {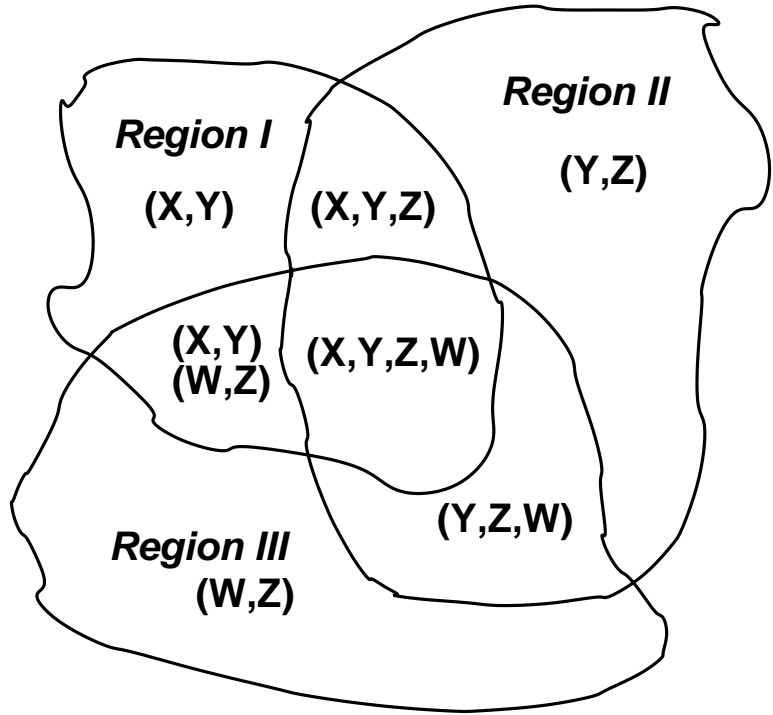}
  }
\caption{A sketch of a landscape in which different regions exhibit
   different correlations between phenomenological observables $X$, $Y$,
   $Z$, and $W$.  As discussed in the text, 
   the overlaps between these regions can then exhibit
   correlations amongst larger subsets of observables
   or multiple independent correlations involving smaller subsets of observables.
   For example, while each region separately exhibits a correlation amongst
   two observables, the overlap between Regions~I and II exhibits
   a single correlation between three observables  while the overlap between
   Regions~I and III exhibits two independent correlations, each involving
   only two observables. Many other generalizations and geometric configurations
   are possible.}  
\label{figone}
\end{figure}

However, we believe that neither of these of these two options is likely to
represent the true nature of correlations on the string landscape.
Rather, we believe that the true nature of such correlations  
lies somewhere between these two extremes
and is more likely to resemble that shown in Fig.~\ref{figone}.
In Fig.~\ref{figone}, some regions of the landscape exhibit certain correlations
and other regions of the landscape exhibit other correlations.
The number of such distinct regions is likely to be vast,
and many of these regions are also likely to have non-trivial overlaps.
For example, we can imagine that one region might principally correspond to 
perturbative heterotic strings (in which worldsheet symmetries such
as conformal invariance and modular invariance play a decisive role
in producing correlations amongst low-energy observables),
while another region might principally correspond to intersecting D-brane models
(in which decisive roles are instead played by tadpole anomaly constraints).
Of course, it is a na\"\i ve expectation that different correlation-class
regions will correspond neatly to different underlying string construction methods,
and more subtle mappings between construction methodologies and correlation
regions will undoubtedly occur. 
For this reason, it is important that
such regions be defined according to their low-energy 
phenomenological predictions and correlations, not according to their
construction methodologies.  Thus these regions need not be disjoint,
and indeed non-trivial overlaps will occur.

Given such a picture,
the precise nature of correlations at a given point on the landscape
is likely to depend rather sensitively on the location of
that point relative to the boundaries of all  possible nearby regions.
For example, in Fig.~\ref{figone}, we observe that
two phenomenological properties $X$ and $Y$ are correlated 
in Region~I, while $Y$ and $Z$ are correlated in Region~II
and $W$ and $Z$ are correlated in Region~III.
Even though each of these regions exhibits only a single correlation
involving two phenomenological quantities, 
we see that the intersections of these regions
nevertheless exhibit a number of different correlation patterns:
\beqn
   {\rm overlap~I~\&~II:}&&~~~ \hbox{single three-quantity correlation (X,Y,Z)}\nonumber\\ 
   {\rm overlap~II~\&~III:}&&~~~ \hbox{single three-quantity correlation (Y,Z,W)}\nonumber\\ 
   {\rm overlap~I~\&~III:}&&~~~ \hbox{two two-quantity correlations (X,Y) and (Z,W)}\nonumber\\ 
   {\rm overlap~I,~II,~\&~III:}&&~~~ \hbox{single four-quantity correlation (X,Y,Z,W)}~. 
\eeqn

Strictly speaking, such a situation fails to yield a single
correlation which holds across the landscape as a whole.
As such, this situation is one in which it might be claimed
that string theory as a whole is non-predictive.
However, even in such a situation,
we can still claim  that string theory is partially predictive
if the sizes of these correlation-class regions are
relatively large compared with the landscape as a whole.
If there exist huge subtracts of the landscape across which 
correlations hold, then we can claim that string theory is entirely
predictive within each such domain.
At the opposite extreme, however, it may turn out that the 
fundamental regions across which such correlations hold
are relatively small.   For example, one could imagine a situation
in which each region is so small that it contains no more than a single
model.
In such a case, we would then claim that string theory is entirely
non-predictive.

In the remainder of this paper, we would like to attach a quantitative
measure to this notion of predictivity.
Specifically, given a situation such as that sketched in Fig.~\ref{figone},
we would like to develop a mathematical measure
of our power to observe correlations on the landscape
and extract some measure of predictivity.

While there are many ways to develop such a mathematical
model, we shall proceed as follows.
At a practical level, we can imagine that we have 
sampled a certain number $x\gg 1$ of models, randomly selected 
across the landscape as a whole.\footnote{In stating that these models
    are selected randomly, we are disregarding the critical issue
    that arises due to the fact that our sampling techniques will
    inevitably introduce biases that distort the apparent space
    of models in non-trivial ways.  Methods of overcoming these
    difficulties were developed in Ref.~\cite{dienes2}, and we shall
    assume in the remainder of this paper that such methods have
    already been utilized and all such distortions have been 
    eliminated as far as possible.} 
Let us assume that we have analyzed the physical observables
predicted from these $x$ models, and we have not observed
any correlations that hold across this set of models.
Clearly, this means that not all $x$ of our models
come from the same region;  at least one model must originate
from a different region.

\begin{figure}
\centerline{
   \epsfxsize 5.0 truein \epsfbox {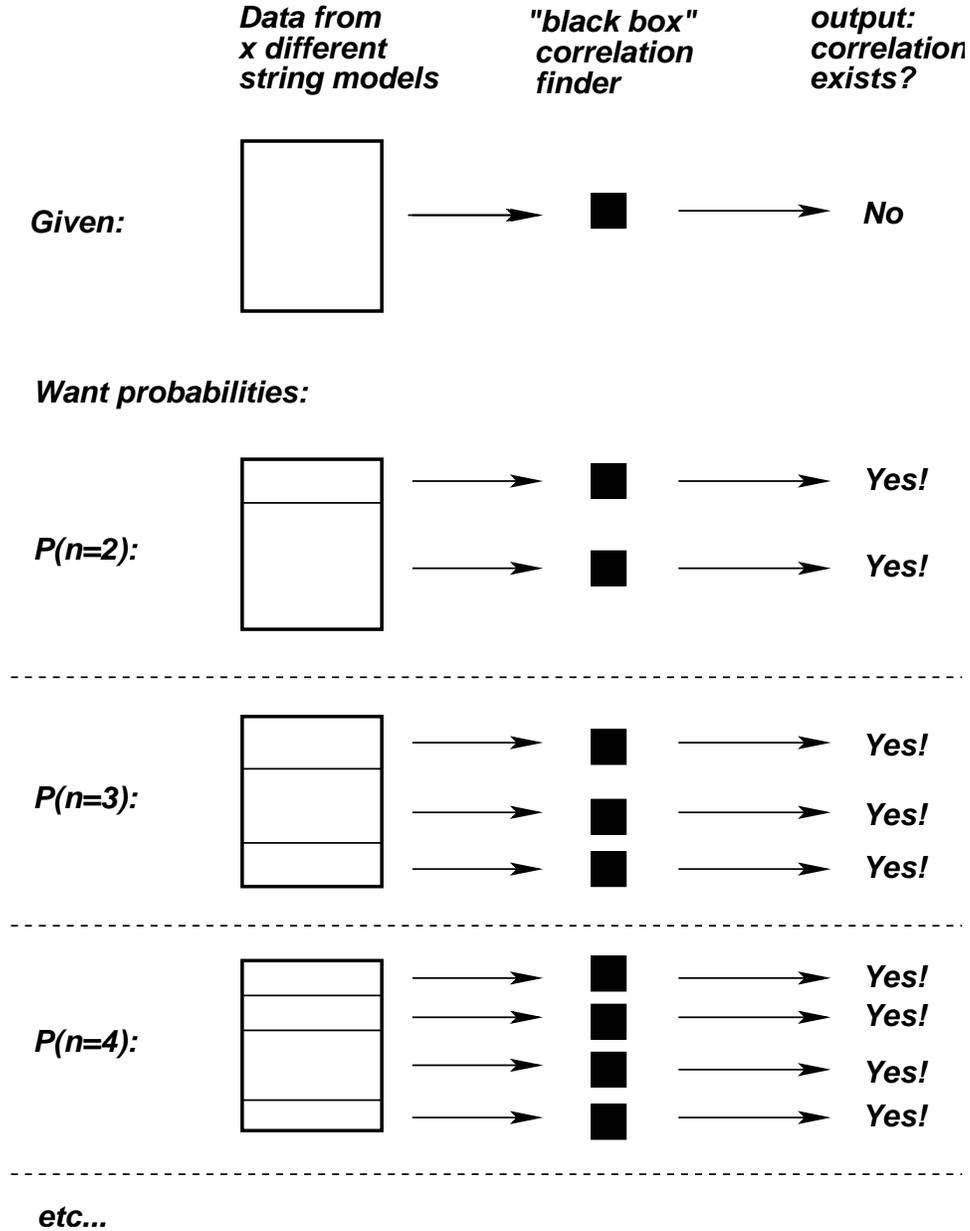}
    }
\smallskip
\caption{Schematic illustration of the fundamental problem.  Suppose data from $x$~string
   models does not exhibit any correlations amongst low-energy physical observables
   which hold across all $x$ models.  What is the probability $P_x(n)$ that we can partition
   our $x$ models into $n$ distinct classes, each of which individually 
   exhibits correlations across the class as a whole?  Clearly $P_x(n)$ grows
   as a function of $n$, ultimately reaching $P_x(n)=1$ for $n=x$ ({\it i.e.}\/, the
   case in which each class is no larger than a single model).  The behavior of $P_x(n)$
   as a function of $n$ for $1<n<x$ determines the extent to which the landscape 
   sketched in Fig.~\protect\ref{figone}
   is predictive, with larger $P_x(n)$ for small $n$ indicating a larger degree
   of predictivity.  }  
\label{figtwo}
\end{figure}

We can then ask for the probability that there exists 
a {\it partitioning}\/ of our data set into two groups of models
such that there exist correlations which hold across each
group separately.  
If no such two-way partitions exist, 
we could then attempt to construct three-way partitions which 
have the same property, and so forth.
In general, we can seek to derive the probability $P_x(n)$ that we can partition
our $x$ models into $n$ distinct classes, each of which individually 
exhibits correlations across the class as a whole.
This question is sketched schematically in Fig.~\ref{figtwo}.

We can immediately make a number of statements concerning $P_x(n)$.
First, $P_x(n)$ will clearly grow monotonically as a function of $n$. 
This follows from the observation that if a given set of $x$ models can be successfully partitioned
into $n$ correlation classes, then it can necessarily be successfully partitioned
into any greater number of correlation classes. 
Second, we observe that $P_x(n)$ should ultimately 
reach $P_x(n)=1$ for $n=x$.
This corresponds to the case 
in which each correlation class is no larger than a single model  ---
although relatively useless, such a partition is indeed guaranteed to be successful. 
Finally, we are intrinsically assuming that $P_x(1)=0$. 
This essentially serves as an initial condition. 

What interests us, however, is the {\it behavior}\/ of 
$P_x(n)$ as a function of $n$ for $1<n<x$, as this determines the extent to which the 
landscape sketched in Fig.~\protect\ref{figone} is predictive. 
Indeed, larger values of $P_x(n)$ for small $n$ can be associated with
a larger degree of predictivity for the landscape as a whole,
in the sense that our correlation classes on the landscape are
larger rather than smaller.

It is important to reiterate that we are defining our correlation
classes of models in terms of their spacetime phenomenological predictions 
rather than their underlying worldsheet or D-brane constructions.  
Needless to say, it is only in this manner that we can declare two
different models to be phenomenologically distinct.
But at a deeper level, we observe that this method of defining
our correlation classes overcomes whatever theoretical prejudices
we might have concerning which phenomenological properties are associated
with which model-construction techniques.
Indeed, one might argue that
the very notion of string theory
being predictive rests on the existence of correlation classes
which {\it transcend}\/ the somewhat artificial boundaries associated
with different underlying model-construction methods.  

We also stress that in this paper, we shall not be concerned
with the inner workings of the ``correlation finder'' sketched
in Fig.~\ref{figtwo}.  Likewise, we shall not be concerned with
the question of {\it how}\/ to partition our $x$ models 
into the $n$ test classes which are then individually examined for internal 
correlations.  Needless to say, these are very important questions ---
the former is critical for data analysis in general, and 
the latter might potentially be addressed through direct enumeration
of different partitioning possibilities or on the basis of other 
external physical information.
However, our purpose in this paper is to study the mathematical extent to
which we can learn about the properties of the underlying landscape,
assuming that such data-analysis tools are at our disposal.

We shall now calculate the probabilities $P_x(n)$.
In order to do so, we shall first need to quantify 
the sizes and overlaps between the correlation regions sketched in Fig.~\ref{figone}.
Let us therefore assume that a given randomly selected string model 
has a probability $p_i$ of being a member of the $i^{\rm th}$ correlation class.
In some sense, the $p_i$ quantify the ``sizes'' of the individual
correlation-class regions across the string landscape.
We shall also need to quantify the sizes of two-region overlaps,
three-region overlaps, and so forth.
Towards this end, we shall let $p_{ij}$ denote the probability
that a randomly selected string model is simultaneously a member of both the
$i^{\rm th}$ and $j^{\rm th}$ correlation classes (where $i\not=j$),
$p_{ijk}$ denote the probability that such a string model 
is simultaneously a member of the $i^{\rm th}$, $j^{\rm th}$, and $k^{\rm th}$ 
correlation classes (where $i$, $j$, and $k$ are all unequal), and so forth.

In general, these quantities $p_{ijk...}$ can vary significantly across
the landscape.  However, for the purposes of calculating the overall
probabilities $P_x(n)$, what really concern us are the ``average'' values
of these quantities.  We shall therefore assume a uniform 
``average'' distribution in which 
\beq 
             p_{i_1,i_2,...,i_N}~=~ a_N \, p~
\label{distribution}
\eeq
where $p$ is an overall arbitrary probability, and where the $a$-coefficients 
satisfy the constraints
\beq
               0~\leq~ ...~\leq~a_4~\leq ~a_3 ~\leq ~a_2 ~\leq ~1~
\label{distribution2}
\eeq
with $a_1\equiv 1$.
There is also another constraint on the $a$-coefficients which will be explained
shortly.

In order to understand these assumptions, it will help to consider 
an abstract geometric picture of the landscape
in which each string model
occupies a volume of arbitrary dimensionality but fixed, uniform magnitude.
We shall refer to the entire space of models arranged this way as the ``correlation space''.
Note that the correlation space is {\it not}\/ the usual geometric picture of the landscape in which
the different directions might be parametrized by different low-energy observables,
or alternatively by different string-construction parameters (\eg, fluxes). 
Indeed, in such a picture, models which are in the same correlation classes
can be scattered across the landscape and need not occupy contiguous regions.
By contrast, in the correlation space, each model occupies an equal volume
of arbitrary (irrelevant) dimensionality,
and models can be freely repositioned so that models in the same correlation
class (according to their low-energy observables) occupy neighboring
contiguous regions, as in Fig.~\ref{figone}.

In terms of the correlation space,
our probability distributions can be understood geometrically as follows.
If we imagine the entire correlation space to occupy a normalized
volume $V=1$, then $p_i$ is nothing but the volume of the $i^{\rm th}$ correlation
region, $p_{ij}$ is nothing but the volume of the $(i,j)$ overlap region,
and so forth.  Likewise, our assumptions in 
Eqs.~(\ref{distribution}) and (\ref{distribution2})
indicate that $p$ is the average volume of 
each correlation class individually, while 
$a_n p$ is the average volume of each overlap region 
between $n$ different correlation classes.

Note that the volume of each overlap region must scale linearly 
with $p$ (the volume of each individual region) 
because our overlap regions will generally have the same 
dimensionalities in the correlation space as each individual region.  
This explains the assumption in Eq.~(\ref{distribution}).
Indeed, this is the major advantage of working with the correlation space
rather than the usual
geometric visualization of the landscape 
in which  models are placed along axes parametrized
by low-energy observables.
In the usual visualization, we would easily expect
situations in which our different correlation classes
have intersections of reduced dimensionalities.  
By contrast, all such situations are automatically incorporated within
the correlation space without any required changes in dimensionality.  

Likewise, the constraint in Eq.~(\ref{distribution2}) merely 
assures that the volume of the average overlap region between $n$ different
correlation classes in the correlation space cannot exceed the volume of the average 
overlap region between $(n-1)$ different correlation classes.
This too makes intuitive sense, since the $n$-overlap region is by definition
more restrictive than the $(n-1)$-overlap region.
Note that the limiting case with $a_2=0$ corresponds to the situation
in which all correlation regions are necessarily disjoint, 
while the case with $a_2=1$ represents a null limit in which
all correlation regions overlap completely.  This implies that $a_3=a_4=...=1$ as well, 
which in turn implies that there is really only
one correlation region.  This implies that $p=1$.

Given the distributions in Eqs.~(\ref{distribution}) and (\ref{distribution2}),
the next step is to calculate the probability $\phi_n$ that a randomly
selected string model is a member of any of $n$ previously selected correlation
classes.  For example, the probability $\phi_1$ that a given model
is a member of a single previously specified correlation class $i$ is nothing but
\beq
         \phi_1~=~ p_i~= p~,
\label{vol1}
\eeq
while the probability $\phi_2$ that a given model is a member of at least one of two
previously specified correlation classes $(i,j)$ is given by
\beq
       \phi_2 ~=~  p_i + p_j - p_{ij} ~=~ 2 p - a_2 p ~=~ (2-a_2)\,p~
\label{vol2}
\eeq
and the  probability $\phi_3$ that a given model is a member of at least one of three
previously specified correlation classes $(i,j,k)$ is given by
\beqn
            \phi_3 &=&  p_i + p_j + p_k - p_{ij} - p_{jk} - p_{ik} + p_{ijk} \nonumber\\
             &=&  3 p - 3 a_2 p + a_3 p \nonumber\\
           &=& (3-3a_2+a_3)\,p~.
\label{vol3}
\eeqn
Note that in the correlation space, each of these results has a natural geometric
interpretation:  $\phi_1$ is the volume of  a single correlation region;
$\phi_2$ is the combined volume of two correlation regions [which is the sum
of the volume of each region minus the (double-counted) volume of their overlap];
and so forth.
In general, for $\ell$ previously specified correlation classes $(i,j,k,r,...,s)$,
we have
\beqn
         \phi_\ell &=&  \sum_i p_i - \sum_{ij} p_{ij} + \sum_{ijk} p_{ijk}
             - \sum_{ijkr} p_{ijkr} +...+ p_{ijkr...s} \nonumber\\
             &=& \left[ \sum_{m=1}^\ell (-1)^{m+1} \,{\ell!\over m!(\ell-m)!}\, 
            a_m\right]\, p~   
\label{volgen}
\eeqn
where the summations in the first line of Eq.~(\ref{volgen}) 
are over all unequal choices from amongst the classes $(i,j,k,r,...,s)$.

Of course, logical consistency requires that $\phi_1\leq \phi_2\leq \phi_3 \leq ...$.
This in turn places an additional constraint on the 
$a$-coefficients in Eq.~(\ref{distribution}).
Thus, while Eq.~(\ref{distribution2}) indicates that each $a_i$ cannot exceed
$a_{i-1}$, we now see that each $a_i$ also cannot be too much smaller than $a_{i-1}$.
This new constraint merely reflects the mathematical fact that if all two-region 
overlaps are large, there is no way to prevent three-region overlaps from also being
fairly large, and so forth.
For example, while we have $a_3\leq a_2$, the requirement that $\phi_3\geq \phi_2$
also requires that $a_3\geq 2a_2-1$.

Given these results for $\phi_\ell$, we now have all of the ingredients necessary
to calculate $P_x(n)$.
Let us begin by calculating the {\it exclusive}\/ probabilities $\hat P_x(n)$ and
see how they evolve as we examine more and more models in the landscape.
Unlike the general probabilities $P_x(n)$ 
that $x$ models will exhibit {\it at least}\/ $n$ different correlation classes,
the exclusive probabilities $\hat P_x(n)$ 
represent the probabilities that 
$x$ models will exhibit {\it exactly}\/ $n$ correlation classes.

When $x=1$, there is only one model and consequently only one correlation
class needed.  We therefore have $\hat P_1(1)=1$ and $\hat P_1(n)=0$ for all $n>1$.
Next, when we select our second model, there are two possibilities:
either it is in the same correlation class as our first model 
(which happens with probability $\phi_1$), or it is not.
We thus find that $\hat P_2(1)=\phi_1=p$, while $\hat P_2(2)=1-\phi_1=1-p$.
Proceeding to the third model, we again have the same situation:
it may be in the same correlation classes as we have already seen,
or it may not.  Tallying the possibilities in each case,
we then find $\hat P_3(1)=\phi_1^2 = p^2$, 
while 
$\hat P_3(2)= \phi_1 (1-\phi_1) + (1-\phi_1)\phi_2= (3-a_2)(p-p^2)$
and 
$\hat P_3(3)= (1-\phi_1)(1-\phi_2)=1-(3-a_2)p + (2-a_2)p^2 $. 

This process continues as we select more and more models.
Ultimately, all of our exclusive probabilities $\hat P_x(n)$ can be 
generated through the recursion relation 
\beq
           \hat P_\ell (k) ~=~ \hat P_{\ell-1}(k) \phi_{k} 
           ~+~ \hat P_{\ell-1}(k-1) \left[ 1-\phi_{k-1}\right]~
\label{recurs}
\eeq
with the initial condition $\hat P_1(1)=1$. 
This recursion relation merely says that there are only two possible
ways of finding $k$ correlation classes after $\ell$ models have 
been examined:  
either there were already $k$ classes found 
from amongst the previous $\ell-1$ models
(and the $\ell^{\rm th}$ model
must be in one of these $k$ classes), or there were only $k-1$ classes
found 
from amongst the previous $\ell-1$ models
(and the $\ell^{\rm th}$ model is not in one of those classes).
These possibilities then give rise to the first and second terms 
on the right side of Eq.~(\ref{recurs}).

Given the recursion relation in Eq.~(\ref{recurs}), 
we immediately see that $\hat P_x(1)=\phi_1^x= p^x$, which is the probability
that $x$ models are all in the same correlation class.  Likewise, we see
that $\hat P_x(x)=\prod_{i=1}^x (1-\phi_i)$, which is the probability that
each successive model is outside the correlation classes determined by the
previous models.  

Finally, given the exclusive probabilities $\hat P_x(n)$, we can easily
calculate the general probabilities $P_x(n)$:
\beq
            P_x(n) ~=~ \sum_{m=1}^n \hat P_x(m)~.
\label{unrestrict}
\eeq
It then follows, for example, that while $P_x(1)=\hat P_x(1)=p^x$,
we have $P_x(x)=1$, as required.

\begin{figure}[b!]
\centerline{
   \epsfxsize 4.0 truein \epsfbox {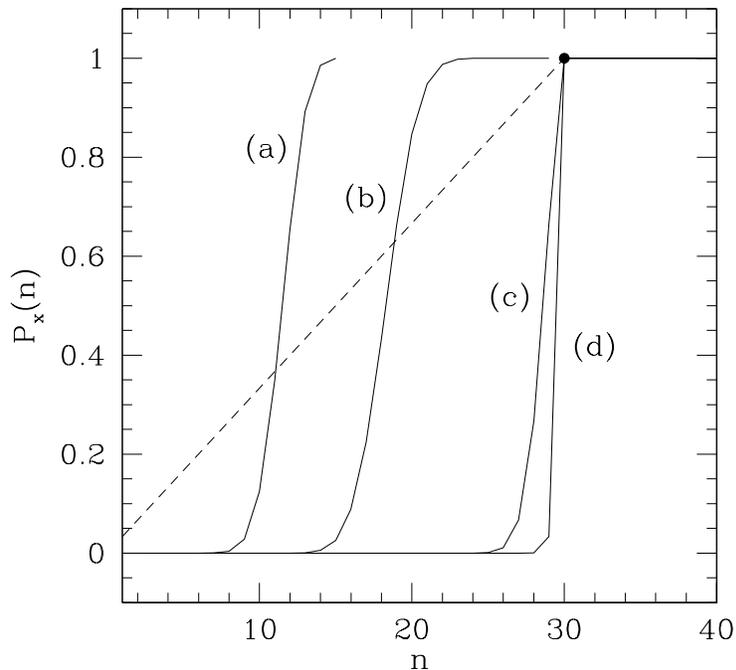}
    }
\smallskip
\caption{The probabilities $P_x(n)$, plotted (solid lines) 
      as functions of $n$ in the range $1\leq n\leq x$ for
   (a) $x=15$, (b) $x=29$, (c) $x=100$, and (d) $x=200$.
   In each case, we have chosen $p=1/30$ and $a_i=0$ for all $i\geq 2$,
   so that our correlation classes are all non-overlapping (disjoint).
   The dashed line shows $\phi_n$ as a function of $n$. 
   For $x\lsim 1/p$, we see that $P_x(n)$ reaches $1$  when $n=x$;
   by contrast, for $x \gg 1/p$, we see that $P_x(n)$ reaches $1$
   near $n\approx 1/p$.  As $x\to\infty$, the curve $P_x(n)$ asymptotes
   to a sharp step function at $n=1/p$.  Thus, as the number of models
   examined increases beyond $1/p$, measuring $P_x(n)$ can yield an extremely
   precise measure for the average value of $p_i$ on the string landscape. 
   }  
\label{plot1}
\end{figure}

Using Eqs.~(\ref{volgen}),
(\ref{recurs}), and (\ref{unrestrict}),
it is straightforward to evaluate $P_x(n)$ as a function of $n$ in the
range $1\leq n \leq x$ for
any $\lbrace p, a_2, a_3, ...\rbrace$.
Our results are shown in Fig.~\ref{plot1} for the case
with $p=1/30$ and $a_i=0$ for all $i\geq 2$,
corresponding to a situation in which there are $30$ disjoint
correlation classes.
Already, we can observe certain general features. 
For $x\lsim 1/p$, we see that $P_x(n)$ reaches $1$ when $n=x$,
as required.  However, for $x \gg 1/p$, 
we see that $P_x(n)$ reaches $1$ near $n\approx 1/p$.  
This too makes sense, since we expect to achieve a successful partition
of our data set when
the number of partitions is approximately equal to $1/p$, the number of 
disjoint correlation classes.
Finally, we observe that as $x\to\infty$, the curve $P_x(n)$ asymptotes
to a sharp step function at $n=n_\ast$ where $n_\ast\equiv [1/p]+1$, \ie,
where $n_\ast$ is the smallest integer exceeding $1/p$. 
This sharpening into a step function also makes intuitive sense.
As we examine more and more models, it becomes more and more unlikely
that we have missed finding at least one representative model from
any correlation class.  Thus, we can only achieve successful partitionings
when the number of partitions equals the number of correlation classes.

This last result provides us with a clear ``experimental'' way of determining
the average value of $p_i$ on the string landscape. 
Indeed, as the number of models increases beyond $1/p$ [which can be
determined from the increasing sharpness of the rise of $P_x(n)$], 
the location of this rise in $P_x(n)$ will be given by $n_\ast$, the 
smallest integer exceeding $1/p$.

These results are valid for the situation in which all correlation classes
are disjoint.  However, 
this general situation persists even 
when the $a$-coefficients are non-zero and overlaps between regions become
significant.
Indeed, with non-zero overlap regions, the volumes
$\phi_n$ will no longer grow linearly with $n$;  
these volumes will accrue more slowly
as a function of $n$ 
because only part of the volume corresponding to each 
new correlation class leads to new territory not previously covered. 
Nevertheless, the previous behavior for $P_x(n)$ persists, provided 
we more generally define $n_\ast$ as the smallest integer $n$ 
for which $\phi_n=1$.  Indeed, just as in the disjoint-region case,
we find that $P_x(n)$ reaches $1$ when $n=x$
for $x\lsim n_\ast$, 
while $P_x(n)$ reaches $1$ near $n\approx n_\ast$
for $x \gg n_\ast$.
Indeed, as $x\to\infty$, the curve $P_x(n)$ continues to asymptote
to a sharp step function at $n=n_\ast$.

\begin{figure}
\vskip -0.3 truein
\centerline{
   \epsfxsize 3.0 truein \epsfbox {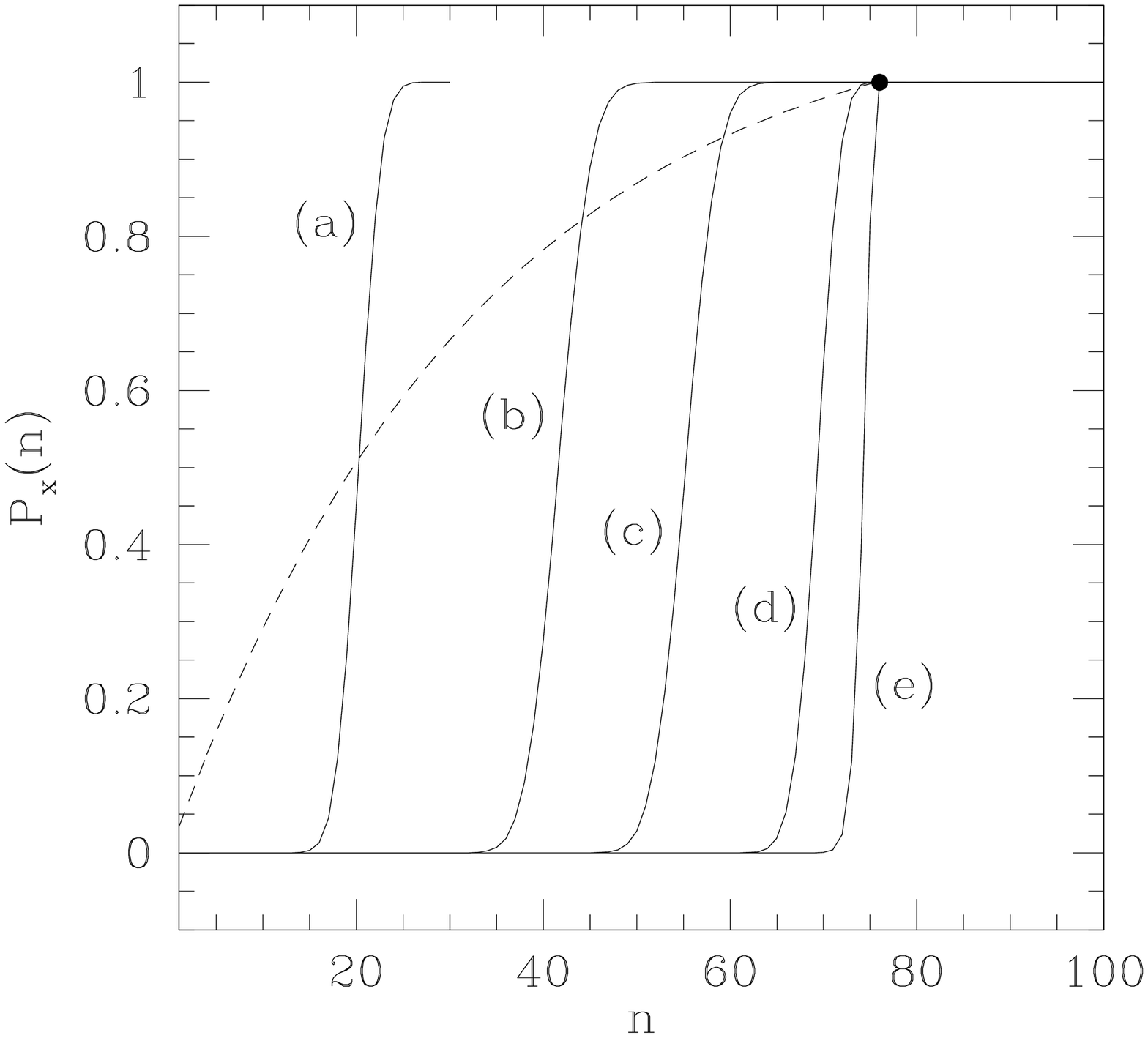}
    }
\vskip -0.3 truein
\caption{The probabilities $P_x(n)$, plotted 
      as functions of $n$ in the range $1\leq n\leq x$ for
   (a) $x=30$, (b) $x=100$, (c) $x=200$, (d) $x=500$, and (e) $x=1000$.
   In each case, we have chosen $p=1/30$.
   However, unlike the plot in Fig.~\protect\ref{plot1}, 
   we have taken $a_n=r^{n-1}$ with $r=0.03$ for all $n\geq 2$, 
   reflecting significant overlaps between correlation-class regions.
   The dashed line shows $\phi_n$ as a function of $n$, reaching $\phi_n=1$  
   at $n_\ast=76$.
   We see that $P_x(n)$ behaves similarly to the case in
   Fig.~\protect\ref{plot1}, 
   with the primary difference that
   significantly larger values of $x$ are required in order to ``saturate'' 
   the probability function and trigger the transition to a step function. 
   Despite these differences, however, 
   we see that measuring $P_x(n)$ 
   for $x\gg n_\ast$ continues to yield an extremely
   precise measure for $n_\ast$ on the string landscape.  
   }  
\label{plot2}
\bigskip
\centerline{
   \epsfxsize 3.0 truein \epsfbox {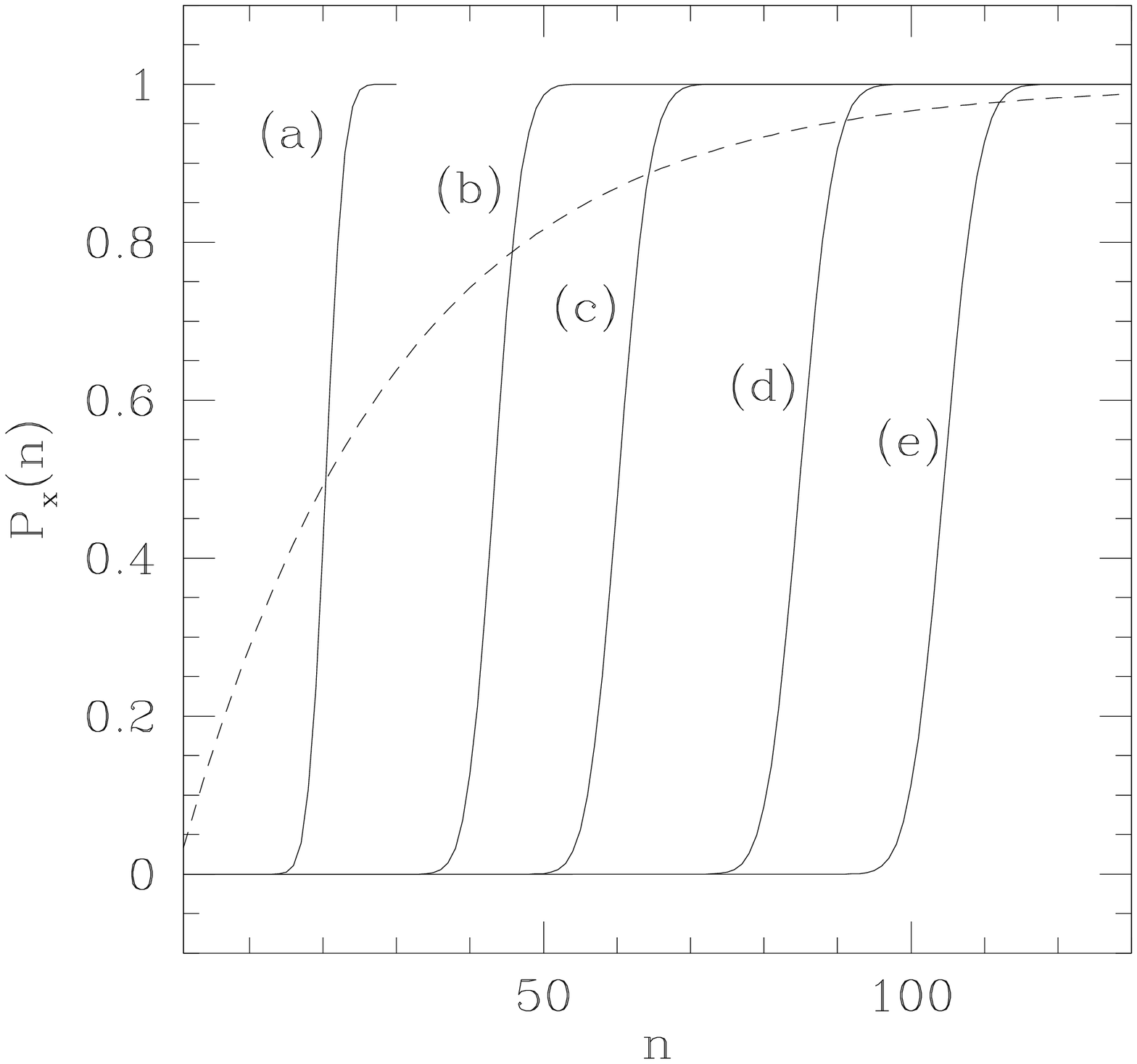}
    }
\vskip -0.3 truein
\caption{The probabilities $P_x(n)$, plotted 
      as functions of $n$ in the range $1\leq n\leq x$ for
   (a) $x=30$, (b) $x=100$, (c) $x=200$, (d) $x=500$, and (e) $x=1000$.
    This plot is the same as in Fig.~\protect\ref{plot2} except that
    we have now taken $r=1/30$.
   As is evident, this change in the value of $r$ (adjusting
    its value by a mere few parts in a thousand) has changed the behavior
   of $P_x(n)$ significantly, shifting $n_\ast \to\infty$ and
    entirely eliminating the asymptotic step-function behavior
    for $P_x(n)$ no matter how large $x$ becomes.
   As argued in the text, this represents a highly fine-tuned
    situation in which the landscape consists of an infinite number
    models and an infinite number of correlation classes.  In such
    a case, string theory would have no predictive power.
   }  
\label{plot3}
\end{figure}


This situation is illustrated in Fig.~\ref{plot2}.
For this figure, we have taken $a_1=1$ and $a_n=r^{n-1}$
where $r$ is a pre-determined scale factor;
note that such $a$-coefficients satisfy all of the 
self-consistency constraints previously discussed.
Also note that even though $\phi_n$ is growing only 
very slowly as a function of $n$, the probabilities $P_x(n)$
still make a relatively sharp transition from $0$ to $1$,
even for $x\leq n_\ast$.
A similar situation emerges for any $r<p$.

Thus, even when there are significant overlaps between correlation
regions on the landscape, we see that we can continue
to extract sharp ``experimental'' data about the landscape merely by
taking $x\gg n_\ast$.  
Indeed, the only difference relative to the disjoint-region case
is that we are now extracting information
about $n_\ast$ rather than about $1/p$.

There is only one finely-tuned situation in which this
method of measuring $P_x(n)$ fails to yield clear information
about the underlying landscape:  this occurs if $n_\ast$
is infinite.
At first glance, it may seem that one cannot ever physically realize
a situation in which $n_\ast$ is infinite.
However, it is possible for $\phi_n$ to approach $1$ as an {\it asymptote}\/
rather than actually hit $1$ for finite $n$.
Again considering the case with $a_n= r^{n-1}$ for all $n\geq 2$,
it turns out that we can mathematically realize such a situation 
by taking $r=p$.  Such a situation is illustrated in Fig.~\ref{plot3},
where we see that our probability function $P_x(n)$ 
fails to reach a fixed shape no matter how large $x$ becomes.

Physically, taking $r=p$ corresponds to a situation 
in which each new correlation class adds an incrementally 
smaller amount of new volume, so that an infinite number of correlation 
classes are required to saturate the full correlation space.
Clearly, such a situation is highly fine-tuned, requiring a landscape
exhibiting both an infinite number of models and an infinite number of correlation
classes.  String theory would have absolutely no predictive power in such
a situation.  However, there exist general arguments~\cite{DouglasTaylor} suggesting 
that the number of string models in the landscape is actually finite.
If so, then such a situation cannot arise.

Likewise, for mathematical completeness, we remark that a similar
situation with infinite $n_\ast$ can also arise in our example by 
taking $r>p$.  In such cases, as $n\to\infty$, the function $\phi_n$ asymptotes
to a value {\it less}\/ than $1$, once again implying that $n_\ast$ is infinite. 
However, this situation is also clearly unphysical, since it corresponds to 
the self-contradictory claim that there exist non-vanishing regions of the landscape 
which are not populated by any string models!

We conclude, then, that measuring $P_x(n)$ provides a robust
practical method of extracting information concerning 
the average behavior of the different correlation classes across
the string landscape. 
This in turn provides a direct and compelling way of
quantifying the extent to which string theory is predictive.
Perhaps the primary virtue of this method is that it can readily
be applied for situations in which only a relatively small number
of string models are examined, provided these models are randomly
selected from across the entire landscape as a whole.
Indeed, all that is required is that $x$, the number of models
examined, exceed $n_\ast$ by perhaps one or two orders of magnitude,
a proposition which can be verified (without {\it a priori}\/ knowledge
of $n_\ast$) by measuring $P_x(n)$ for increasing values of $x$ and observing
if and when this function saturates into step-function behavior.

Needless to say, the calculations in this paper may be easily 
generalized to more complex landscape distributions and correlation-region
overlap patterns.  
However, the central point of this paper is general and remains applicable
regardless of such possible generalizations:  there will always be
a value $n_\ast$ at which $\phi_n=1$, and this value can be 
``experimentally'' extracted with great statistical certainty
through the methods we have described.
Indeed, we have shown this explicitly for landscape distributions
at both extremes:  distributions in which our correlation-class 
regions are entirely disjoint, and distributions in which significant overlaps occur.

Note that even our notion of ``correlation class''
can be generalized without altering the main results of
this paper.  In this paper, we have implicitly assumed that
within a single correlation class, there exists a tight mathematical relation
between specific low-energy observables.  However, 
this requirement may also be relaxed:  meaningful correlation classes may
also exist in which one might be able to say nothing more than that a certain
 {\it range}\/ of values for one specific low-energy observable
tends to be statistically correlated
with a certain {\it range}\/ of values for a different low-energy observable.
Indeed, evidence that such types of correlation classes
exist has recently been presented in Ref.~\cite{gordy}.   
Neverthless, the methods we have developed in this paper are
applicable to these types of generalized correlation classes as well.

It is perhaps premature to speculate about a likely value of $n_\ast$
across the string landscape, but we would imagine that $n_\ast$ should
not exceed ${\cal O}(10)$ at most if string theory is to have any
meaningful predictive power.  As we have discussed, such correlation 
classes might correspond,
for example, to different types of model construction, or different
topological classes of compactification geometry.  
In such cases, obtaining and analyzing a sufficient number of
string models should not be difficult.

Of course, our method of examining $P_x(n)$ can also
be used to examine the properties of any {\it subset}\/ of the landscape.
For example, one might restrict to a class of models
which share a common underlying construction methodology.
In such cases, the resulting information for $n_\ast$ then applies
to the correlation regions appropriate for that subset.
Our method is therefore suitable for examinations of arbitrary 
subsets of the landscape, without requiring knowledge of the
string landscape as a whole.

\bigskip

\section*{Acknowledgments}

We wish to thank the CERN Theory Division, where
this work was initiated, for hospitality during the 
July/August 2008 TH Institute on String Phenomenology. 
The work of KRD was supported in part by the
U.S. Department of Energy under Grant~DE-FG02-04ER-41298
and by the
U.S. National Science Foundation under Grant PHY/0301998.
The work of ML was supported in part
by ANR grant ANR-05-BLAN-0079-02,
RTN contracts MRTN-CT-2004-005104 and
MRTN-CT-2004-503369, CNRS PICS~\#2530, 3059 and 3747,
and by a European Union Excellence Grant MEXT-CT-2003-509661.
We are happy to thank W.~Taylor for discussions.

\bibliographystyle{unsrt}

\begin{thebibliography}{99}


\bibitem{landscape}
  S.~Kachru, R.~Kallosh, A.~Linde and S.~P.~Trivedi,
  Phys.\ Rev.\ D {\bf 68}, 046005 (2003)
  [arXiv:hep-th/0301240];\\
  L.~Susskind,
  arXiv:hep-th/0302219.\\
   For popular introductions, see:\\
   R.~Bousso and J.~Polchinski,
  Sci.\ Am.\  {\bf 291}, 60 (2004);\\
   S.~Weinberg,
  arXiv:hep-th/0511037;\\
  A.~N.~Schellekens,
  arXiv:0807.3249 [physics.pop-ph].




\bibitem{douglas}
  M.~R.~Douglas,
  JHEP {\bf 0305}, 046 (2003)
  [arXiv:hep-th/0303194].

\bibitem{abstract2}
  S.~Ashok and M.~R.~Douglas,
  JHEP {\bf 0401}, 060 (2004)
  [arXiv:hep-th/0307049];\\
  F.~Denef and M.~R.~Douglas,
  JHEP {\bf 0405}, 072 (2004)
  [arXiv:hep-th/0404116];\\
  A.~Giryavets, S.~Kachru and P.~K.~Tripathy,
  JHEP {\bf 0408}, 002 (2004)
  [arXiv:hep-th/0404243];\\
   A.~Misra and A.~Nanda,
  Fortsch.\ Phys.\  {\bf 53}, 246 (2005)
  [arXiv:hep-th/0407252];\\
  M.~R.~Douglas,
  Comptes Rendus Physique {\bf 5}, 965 (2004)
  [arXiv:hep-th/0409207];\\
  J.~Kumar and J.~D.~Wells,
  Phys.\ Rev.\ D {\bf 71}, 026009 (2005)
  [arXiv:hep-th/0409218];
  JHEP {\bf 0509}, 067 (2005)
  [arXiv:hep-th/0506252];
  arXiv:hep-th/0604203;\\
  F.~Denef and M.~R.~Douglas,
  JHEP {\bf 0503}, 061 (2005)
  [arXiv:hep-th/0411183];\\
  O.~DeWolfe, A.~Giryavets, S.~Kachru and W.~Taylor,
  JHEP {\bf 0502}, 037 (2005)
  [arXiv:hep-th/0411061];\\
  B.~S.~Acharya, F.~Denef and R.~Valandro,
  JHEP {\bf 0506}, 056 (2005)
  [arXiv:hep-th/0502060];\\
  F.~Denef and M.~R.~Douglas,
  arXiv:hep-th/0602072;\\
   J.~Shelton, W.~Taylor and B.~Wecht,
  arXiv:hep-th/0607015;\\
   A.~Hebecker and J.~March-Russell,
  arXiv:hep-th/0607120.

\bibitem{susybreakingabstract}
  M.~R.~Douglas,
  arXiv:hep-th/0405279;\\
  M.~Dine, E.~Gorbatov and S.~D.~Thomas,
  arXiv:hep-th/0407043;\\
  M.~Dine, D.~O'Neil and Z.~Sun,
  JHEP {\bf 0507}, 014 (2005)
  [arXiv:hep-th/0501214];\\
  JHEP {\bf 0601}, 162 (2006)
  [arXiv:hep-th/0505202];\\
  M.~Dine and Z.~Sun,
  JHEP {\bf 0601}, 129 (2006)
  [arXiv:hep-th/0506246].



\bibitem{DouglasTaylor}
  M.~R.~Douglas and W.~Taylor,
  arXiv:hep-th/0606109;\\
  B.~S.~Acharya and M.~R.~Douglas,
  arXiv:hep-th/0606212.


\bibitem{direct}
  R.~Blumenhagen, F.~Gmeiner, G.~Honecker, D.~Lust and T.~Weigand,
  Nucl.\ Phys.\ B {\bf 713}, 83 (2005)
  [arXiv:hep-th/0411173];
  JHEP {\bf 0601}, 004 (2006)
  [arXiv:hep-th/0510170];\\
  F.~Gmeiner, D.~Lust and M.~Stein,
  JHEP {\bf 0705}, 018 (2007)
  [arXiv:hep-th/0703011];\\
  F.~Gmeiner and G.~Honecker,
  JHEP {\bf 0709}, 128 (2007)
  [arXiv:0708.2285 [hep-th]];
  JHEP {\bf 0807}, 052 (2008)
  [arXiv:0806.3039 [hep-th]].\\
  For a review, see:  \\
  F.~Gmeiner,
  arXiv:hep-th/0608227.


\bibitem{direct2}
  T.~P.~T.~Dijkstra, L.~R.~Huiszoon and A.~N.~Schellekens,
  Phys.\ Lett.\ B {\bf 609}, 408 (2005)
  [arXiv:hep-th/0403196];
  Nucl.\ Phys.\ B {\bf 710}, 3 (2005)
  [arXiv:hep-th/0411129];\\
  P.~Anastasopoulos, T.~P.~T.~Dijkstra, E.~Kiritsis and A.~N.~Schellekens,
  Nucl.\ Phys.\  B {\bf 759}, 83 (2006)
  [arXiv:hep-th/0605226].


\bibitem{direct3}
  J.~P.~Conlon and F.~Quevedo,
  JHEP {\bf 0410}, 039 (2004)
  [arXiv:hep-th/0409215];\\
  S.~S.~AbdusSalam, J.~P.~Conlon, F.~Quevedo and K.~Suruliz,
  JHEP {\bf 0712}, 036 (2007)
  [arXiv:0709.0221 [hep-th]].


\bibitem{dienes}
    K.~R.~Dienes,
    Phys.\ Rev.\ D {\bf 73}, 106010 (2006)
    [arXiv:hep-th/0602286].


\bibitem{dienes2}
    K.~R.~Dienes and M.~Lennek,
  Phys.\ Rev.\  D {\bf 75}, 026008 (2007)
  [arXiv:hep-th/0610319].


\bibitem{dienes3}
    K.~R.~Dienes, M.~Lennek, D.~Senechal and V.~Wasnik,
  Phys.\ Rev.\  D {\bf 75}, 126005 (2007)
  [arXiv:0704.1320 [hep-th]];
  New J.\ Phys.\  {\bf 10}, 085003 (2008)
  [arXiv:0804.4718 [hep-ph]].



\bibitem{direct4}
  O.~Lebedev, H.~P.~Nilles, S.~Raby, S.~Ramos-Sanchez, M.~Ratz,
  P.~K.~S.~Vaudrevange and A.~Wingerter,
  Phys.\ Lett.\  B {\bf 645}, 88 (2007)
  [arXiv:hep-th/0611095];\\
  S.~Raby and A.~Wingerter,
  Phys.\ Rev.\  D {\bf 76}, 086006 (2007)
  [arXiv:0706.0217 [hep-th]];\\
  B.~Dundee, S.~Raby and A.~Wingerter,
  arXiv:0805.4186 [hep-th];\\
  O.~Lebedev, H.~P.~Nilles, S.~Ramos-Sanchez, M.~Ratz and
  P.~K.~S.~Vaudrevange,
  arXiv:0807.4384 [hep-th];\\
  B.~Dundee and S.~Raby,
  arXiv:0808.0992 [hep-th].


\bibitem{Faraggi}
   A.~E.~Faraggi, C.~Kounnas and J.~Rizos,
  arXiv:hep-th/0606144;
  arXiv:hep-th/0611251.


\bibitem{direct4a}
  P.~Candelas, X.~de la Ossa, Y.~H.~He and B.~Szendroi,
  arXiv:0706.3134 [hep-th];\\
  M.~Gabella, Y.~H.~He and A.~Lukas,
  arXiv:0808.2142 [hep-th].
  

\bibitem{gordy}
G.~L.~Kane, P.~Kumar and J.~Shao,
  J.\ Phys.\ G {\bf 34}, 1993 (2007)
  [arXiv:hep-ph/0610038];
  Phys.\ Rev.\  D {\bf 77}, 116005 (2008)
  [arXiv:0709.4259 [hep-ph]].


\bibitem{fieldtheory}
  K.~R.~Dienes, E.~Dudas and T.~Gherghetta,
  Phys.\ Rev.\ D {\bf 72}, 026005 (2005)
  [arXiv:hep-th/0412185];\\
  N.~Arkani-Hamed, S.~Dimopoulos and S.~Kachru,
  arXiv:hep-th/0501082;\\
  J.~Distler and U.~Varadarajan,
  arXiv:hep-th/0507090;\\
   B.~Feldstein, L.~J.~Hall and T.~Watari,
  arXiv:hep-ph/0608121;\\
  V.~Balasubramanian, J.~de Boer and A.~Naqvi,
  arXiv:0805.4196 [hep-th].


\bibitem{reviews}
  J.~Kumar,
  arXiv:hep-th/0601053;\\
  M.~R.~Douglas and S.~Kachru,
  arXiv:hep-th/0610102.





\end{thebibliography}

\end{document}